\documentclass[superscriptaddress,secnumarabic,nobibnotes,aps,prd,showkeys,noshowpacs,onecolumn,12pt]{revtex4}
\usepackage{eurosym}
\usepackage{graphics}
\usepackage{graphicx}
\usepackage{color}
\usepackage{epsf}
\usepackage{bm}
\usepackage{amsmath,amssymb,amsfonts,amsthm}
\usepackage{latexsym}
\usepackage{enumerate}
\usepackage{hyperref}

\begin{document}

\title{Cosmological Interactions with Phantom Scalar Field: Revisiting
Background Phase-Space Analysis with Compactified Variables}
\author{Genly Leon}
\email{genly.leon@ucn.cl}
\affiliation{Departamento de Matem\'{a}ticas, Universidad Cat\'{o}lica del Norte, Avda.
Angamos 0610, Casilla 1280 Antofagasta, Chile}
\affiliation{Institute of Systems Science, Durban University of Technology, Durban 4000,
South Africa}
\author{Daya Shankar}
\email{daya.shankar@woxsen.edu.in}
\affiliation{School of Sciences, Woxsen University, Hyderabad 502345, Telangana, India}
\author{Amlan Halder}
\email{amlankanti.halder@woxsen.edu.in}
\affiliation{School of Sciences, Woxsen University, Hyderabad 502345, Telangana, India}
\author{Andronikos Paliathanasis}
\email{anpaliat@phys.uoa.gr}
\affiliation{Institute of Systems Science, Durban University of Technology, Durban 4000,
South Africa}
\affiliation{School for Data Science and Computational Thinking and Department of Mathematical Sciences, Stellenbosch University,
Stellenbosch, 7602, South Africa} 
\affiliation{Departamento de Matem\'{a}ticas, Universidad Cat\'{o}lica del Norte, Avda. Angamos 0610, Casilla 1280 Antofagasta, Chile}
\affiliation{School of Sciences, Woxsen University, Hyderabad 502345, Telangana, India}

\begin{abstract}
Energy transfer in the dark sector of the universe gives rise to new
phenomena of special interest in modern cosmology. When dark energy is modeled as a phantom scalar field, interactions become crucial to avoid Big Rip singularities. In this work, we revisit the
phase-space analysis of the field equations by introducing a new set of
dimensionless variables distinct from the traditional Hubble normalization
approach. These new variables define a compactified phase space for the
evolution of physical parameters. We demonstrate that these compactified variables offer fresh insights into the phase-space analysis in gravitational theories, particularly when the dark energy fluid is allowed
to possess a negative energy density.
\end{abstract}

\maketitle

\section{Introduction}

\label{sec1}

The detailed analysis of cosmological observations \cite%
{Teg, Kowal, Komatsu,suzuki} implies that the universe is dominated by two matter components that do not interact with light: dark matter and dark energy. Dark matter was introduced to explain the velocity profiles of
galaxies and, on cosmic scales, is described as a dust fluid source \cite%
{dm1}. On the other hand, dark energy \cite{de1,de2} has a negative pressure component and is responsible for the current cosmic acceleration. Although the cosmological constant is the simplest dark energy candidate, it cannot explain the entire cosmological history \cite{cc0}. As a result, a plethora of dark energy models have been proposed in the literature \cite%
{ratra,peebles,ph1,kes,nik,gal02,ndim,sa1,sa3,sa8,sa11}.

Gravitational theories are inherently nonlinear, and analytic methods are
essential for a deeper understanding of cosmic evolution and the viability
of proposed models. In the case of a spatially flat Friedmann--Lemaitre--Robertson--Walker (FLRW) universe, introducing the cosmological
constant leads to a set of differential equations that describe the physical
parameters. These equations can be linearized and admit closed-form
solutions. However, in modified theories of gravity or when scalar fields
are introduced, the theory of variational symmetries \cite{ns1,ns2,ns3,ns4}
and singularity analysis \cite{ns5,ns6} have been widely applied to derive
analytic solutions. The Eisenhart-Duval lift \cite{cc01} recently has been
introduced for analyzing cosmological field equations \cite{ns7,ns8}.

Phase-space analysis remains a powerful tool for studying nonlinear
gravitational models. In this approach, the field equations are expressed as dimensionless variables and stationary points are determined. Each
stable/unstable stationary point corresponds to a specific asymptotic late-time/early-time solution, enabling the
identification of cosmological eras and epochs provided by a given
gravitational model. The phase-space analysis for the exponential potential
in the quintessence model was performed in \cite{cop1}, while it was explored in \cite{cop2,cop3} for nonminimally coupled scalar fields. For
the phantom scalar field, dynamical system analysis is presented in \cite%
{cop4}, for the Galileon in \cite{cop5}, and for multiscalar field models in 
\cite{cop6,cop7,cop8,cop9,cop10}. Dynamical system analysis has also been
extensively applied to anisotropic and inhomogeneous gravitational models 
\cite{cop11,cop12,cop13,cop14,cop15}. This analysis is essential for
assessing the viability of gravitational models, introducing constraints on
the free parameters \cite{am1}, and addressing the initial value problem 
\cite{iv1,iv2}.

This study focuses on the phase-space analysis of a cosmological model
with interactions \cite{Amendola:1999er,cc001,cc002,cc003} between dark
energy and dark matter. Cosmological interacting scenarios have been
proposed to address the coincidence problem \cite{con2}, and cosmological
tensions \cite{ht4,ht5,ht7}. We assume that a
phantom scalar field describes dark energy \cite{q15}, allowing the energy density of the scalar
field to be negative and the violation of the weak energy condition.
Phase-space analysis of interacting models with a scalar field has been
examined in \cite{ds1,ds2,ds3,ds4,ds5}. Cosmological interactions involving
a phantom scalar field were investigated in \cite{var00}, where it was found
that a nonzero interaction term can avoid Big Rip singularities. Recently,
new interacting models \cite{sp1} have been introduced where the interaction
function is not proportional to the scalar field's derivative. As detailed
in \cite{sp2}, these models suffer from singularities that make them unviable.

In the following, we introduce a new set of dimensionless variables distinct
from the Hubble normalization approach. We review previous results on the
cosmological interactions between the phantom field and dark matter. The new
set of variables is compactified, allowing for the direct determination of
all stationary points. These variables provide fresh insights into the
analysis of cosmological models. The structure of the paper is as follows:

In Section \ref{sec2}, we introduce the concept of dark matter-dark energy
interaction, where dark matter is described as a pressureless ideal gas and
dark energy as a phantom scalar field. The spatially flat FLRW metric is
adopted for the geometry of space. In Section \ref{sec3}, we introduce the
new dimensionless variables normalized with a positive quantity dependent
on the cosmological fluid. Unlike the Hubble normalization approach, where
the dimensionless variables are not compactified and require a second set
for detailed analysis, these new variables are compactified from the outset.
To illustrate their novelty, we introduce two interacting models, assume an
exponential potential for the scalar field, and perform a detailed analysis
of stationary points and asymptotic solutions. Finally, we present our conclusions in Section \ref{sec4}.

\section{Cosmological Interactions}

\label{sec2}

On very large scales, the universe is described by the line element of the
spatially flat FLRW geometry 
\begin{equation}
ds^{2}=-dt^{2}+a^{2}(t)\left( dx^{2}+dy^{2}+dz^{2}\right) ,  \label{metric1}
\end{equation}%
where $a(t)$ is the expansion scale factor of the universe.

Within the framework of General Relativity with two fluid components, the
phantom field, and the dark matter, the gravitational field equations for
the line element (\ref{metric1}) are%
\begin{equation*}
G_{\mu \nu }=T_{\mu \nu }^{\left( \phi \right) }+T_{\mu \nu }^{\left(
m\right) },
\end{equation*}%
in which $G_{\mu \nu }$ is the Einstein tensor and $T_{\mu \nu }^{\left(
\phi \right) }$, $T_{\mu \nu }^{\left( m\right) }$ are the energy-momentum
tensors for the phantom field and the dark matter defined as
\begin{eqnarray}
T_{\mu \nu }^{\left( \phi \right) } &=&-\phi _{,\mu }\phi _{,\nu }+g^{\mu
\nu }\left( -\frac{1}{2}g^{\kappa \sigma }\phi _{,\kappa }\phi _{,\sigma
}+V\left( \phi \right) \right) , \\
T_{\mu \nu }^{\left( m\right) } &=&\rho _{m}u_{\mu }u_{\nu },
\end{eqnarray}%
in which $u_{\mu }=\delta _{\mu }^{t}$ is the comoving observer, and we have
assumed that the cosmological fluid inherits the symmetries of the
background geometry (\ref{metric1}), that is, $\phi =\phi \left( t\right) $
and $\rho _{m}=\rho _{m}\left( t\right) $. The function $V\left( \phi \right) $ gives the scalar field potential that drives the dynamics.

The cosmological field equations read%
\begin{eqnarray}
3H^{2} &=&\rho _{\phi }+\rho _{m}, \\
-2\dot{H}-3H^{2} &=&p_{\phi },
\end{eqnarray}%
in which 
\begin{align}
\rho _{\phi }& =-\frac{1}{2}\,\dot{\phi}^{2}+V(\phi ),  \label{energy} \\
p_{\phi }& =-\frac{1}{2}\,\dot{\phi}^{2}-V(\phi ),  \label{pressure}
\end{align}%
and $H=\frac{\dot{a}}{a}$ is the Hubble function.

The Bianchi identity leads to the continuous equation 
\begin{equation}
T_{~~~~\ ~~;\nu }^{\left( \phi \right) \mu \nu }+T_{~~~~\ ~~;\nu }^{\left(
m\right) \mu \nu }=0,
\end{equation}%
or equivalently,%
\begin{align}
\dot{\rho}_{m}+3H\rho _{m}& =Q,  \label{mc} \\
\dot{\rho}_{\phi }+3H(\rho _{\phi }+p_{\phi })\rho _{\phi }& =-Q,  \label{fc}
\end{align}%
where function $Q$ defines the interacting scenario.

In the following, we consider the function $Q$ is defined by the following two models \cite{Amendola:2006dg,Pavon:2007gt}: 
\begin{eqnarray*}
\text{Model~}A &:&Q_{A}=\beta _{0}\dot{\phi}\rho _{m}, \\
\text{Model }B &:&Q_{B}=\beta _{0}\dot{\phi}\rho _{\phi }.
\end{eqnarray*}%
Where necessary the interactions are proportional to  $\dot{\phi}$ in
order to avoid the appearance of singularities, see the discussion in \cite%
{sp2}.

At this point, it is important to mention that the interacting model $Q_{A}$
has a geometric origin, and it follows from the Weyl Integrable Spacetime 
\cite{wi1,wi2}, and it describes the Chameleon mechanism \cite{ch1,ch2}.

\section{Compactified Normalization}

\label{sec3} In contrast to the Hubble normalization studied in the recent
work \cite{sp2}; we adopt a different normalization approach based on
matter. Following the methodology established in \cite{ancoley}, we
introduce a new set of dimensionless variables%
\begin{equation}
\chi =\frac{\dot{\phi}}{\sqrt{2}D},~\zeta ^{2}=\frac{V\left( \phi \right) }{%
D^{2}},~\xi ^{2}=\frac{\rho _{m}}{D^{2}},~\eta =\frac{\sqrt{3}H}{D},~\lambda
=\frac{V_{,\phi }}{V},~d\tau =Ddt  \label{cc.01}
\end{equation}%
where%
\begin{equation}
D=\sqrt{\frac{1}{2}\dot{\phi}^{2}+V\left( \phi \right) +\rho _{m}}.
\label{cc.02}
\end{equation}

Furthermore, by definition, it follows%
\begin{equation}
\Xi \equiv \chi ^{2}+\zeta ^{2}+\xi ^{2}=1,  \label{cc.03}
\end{equation}%
which means that variables $\chi ,~\zeta $ and $\xi $ are compactified and
take values on the surface of a unitary sphere.

In terms of the new dimensionless variables (\ref{cc.01}) the first
Friedmann equation reads%
\begin{equation}
-\eta ^{2}-\chi ^{2}+\zeta ^{2}+\xi ^{2}=0.  \label{cc.03a}
\end{equation}%
which provides a second constraint for the dynamical variables.

\subsection{Interacting Model $Q_{A}$}

In terms of the new variables, the field equations for model $Q_{A}$ read%
\begin{eqnarray}
\frac{d\chi }{d\tau } &=&\frac{1}{2}\left( 4\lambda \zeta ^{2}\left( \zeta
^{2}+\xi ^{2}\right) -2\beta _{0}\xi ^{2}-\left( \sqrt{6}\eta \chi \xi
^{2}-4\beta _{0}\xi ^{4}+2\zeta ^{2}\left( \lambda -\beta _{0}\xi ^{2}+\sqrt{%
6}\eta \chi \right) \right) \right) ,  \label{cc.04} \\
\frac{d\zeta }{d\tau } &=&\frac{\zeta }{2}\left( \left( \sqrt{3}\eta \left(
2\chi ^{2}+\xi ^{2}\right) \right) +\left( \chi ^{2}+\xi ^{2}\right) \left( 
\sqrt{2}\left( \lambda -\beta _{0}\xi ^{2}\right) \chi \right) -\zeta
^{2}\left( \sqrt{2}\left( \lambda +a\xi ^{2}\right) \right) \right) ,
\label{cc.05} \\
\frac{d\xi }{d\tau } &=&\frac{\xi }{2}\left( \sqrt{2}\eta \chi \left( \beta
_{0}\zeta ^{4}-2\zeta ^{2}\left( \lambda -\beta _{0}\chi ^{2}\right) +\beta
_{0}\left( \chi ^{4}-\xi ^{4}\right) \right) +\sqrt{3}\left( \chi ^{2}-\zeta
^{2}\right) \right) ,  \label{cc.06} \\
\frac{d\eta }{d\tau } &=&\frac{1}{2}\left( \sqrt{3}\left( 1-\left( \chi
^{2}-\zeta ^{2}\right) +\xi ^{2}\eta ^{2}\right) +2\sqrt{2}\eta \chi \left(
\zeta ^{2}\left( \lambda +\beta _{0}\xi ^{2}\right) +\beta _{0}\xi
^{2}\left( \chi ^{2}+\xi ^{2}\right) \right) \right) ,  \label{cc.07} \\
\frac{d\lambda }{d\tau } &=&\sqrt{2}\lambda \chi \left( \Gamma \left(
\lambda \right) -1\right) ,~\Gamma \left( \lambda \right) =\frac{V_{,\phi
\phi }V}{\left( V_{,\phi }\right) ^{2}}.  \label{cc.08}
\end{eqnarray}

In terms of the new variables, the deceleration parameter reads%
\begin{equation}
q\left( \chi ,\zeta ,\eta \right) =\frac{1}{2}-\frac{3}{2\eta ^{2}}\left(
\chi ^{2}+\zeta ^{2}\right) .  \label{cc.08b}
\end{equation}%
From the latter expression it follows that $q\left( \chi ,\zeta ,\eta
\right) \leq \frac{1}{2}$.

Using the constraint equations (\ref{cc.03}) and (\ref{cc.03a}), and
considering an exponential scalar field potential where $\Gamma \left(
\lambda \right) = 1$, the five-dimensional dynamical system is reduced to a
two-dimensional compactified dynamical system.%
\begin{small}
\begin{align}
\frac{d\chi }{d\tau } &=\frac{\zeta ^{2}}{4}\left( 4\left( \lambda -\beta
_{0}\right) \left( 1-2\chi ^{2}\right) -2\chi \sqrt{6\left( 1-2\chi
^{2}\right) }-\left( 1-\chi ^{2}\right) \left( \chi \left( 4\beta _{0}+\sqrt{%
\left( 1-2\chi ^{2}\right) }\right) \right) -2\beta _{0}\right) ,
\label{cc.10} \\
\frac{d\zeta }{d\tau } &=\frac{1}{2\sqrt{2}}\left( 2\chi \left( \left(
\lambda -2\beta _{0}+2\left( \beta _{0}-\lambda \right) \zeta ^{2}\right)
\chi +2\beta _{0}\chi ^{2}\right) +\sqrt{6}\left( 1-\zeta ^{2}+\chi
^{2}\right) \sqrt{1-2\chi ^{2}}\right) .  \label{cc.11}
\end{align}%
\end{small}
in which we have replaced%
\begin{eqnarray}
\xi &=&\sqrt{1-\chi ^{2}-\zeta ^{2}}, \\
\eta &=&\sqrt{1-2\chi ^{2}},
\end{eqnarray}%
we have decided to work at the branch where $H>0$. We observe that the
system is well defined for $\chi ^{2}\leq \frac{1}{2}$. Furthermore the
field equations (\ref{cc.10}), (\ref{cc.11}) are invariant on the change of
variables $\zeta \rightarrow -\zeta $ and $\xi \rightarrow -\xi $. Without
loss of generality, in the following, we consider the region $\zeta \geq
0,~\xi \geq 0$.

We proceed with the presentation of the stationary points $A=\left( \chi
\left( A\right) ,\zeta \left( A\right) \right) $ of the two-dimensional
dynamical system (\ref{cc.10}), (\ref{cc.11}). The admitted stationary points are

\begin{eqnarray*}
A_{1}^{\pm } &=&\frac{1}{\sqrt{2}}\left( \pm 1,0\right) , \\
A_{2}^{\pm } &=&\frac{1}{\sqrt{2}}\left( \pm 1,1\right) , \\
A_{3} &=&\left( \frac{\lambda }{\sqrt{2\left( 3+\lambda ^{2}\right) }},\sqrt{%
\frac{6+\lambda ^{2}}{6+2\lambda ^{2}}}\right) \\
A_{4}^{\pm } &=&\frac{\sqrt{2}\beta _{0}}{\sqrt{3+4\beta _{0}^{2}}}\left(
\pm 1,0\right) , \\
A_{5} &=&\sqrt{\frac{3}{3+2\left( \alpha -\lambda \right) ^{2}}}%
\left( 1,\sqrt{\frac{2\alpha \left( \alpha -\lambda \right) -3}{3}}\right).
\end{eqnarray*}

The stationary points~$A_{1}^{\pm }$ and $A_{2}^{\pm }$ are defined at the
extreme limits of the variables $\chi $ and $\zeta $. For the physical
parameters $\xi $ and $q$, we calculate the asymptotic solutions at these
points as $\left( \xi ,q\right) _{{A_{1}^{\pm }}}=\left( \frac{1}{\sqrt{2}}%
,-\infty \right) $ and $\left( \xi ,q\right) _{{A_{2}^{\pm }}}=\left(
0,-\infty \right) $.

From these results, we infer that at the asymptotic solutions, corresponding
to $A_{2}^{\pm }$, only the phantom scalar field contributes to the cosmic
fluid and these solutions describe Big Rip singularities.

Analysis of the linearized system near these points reveals that the
asymptotic solutions are always unstable. Specifically, for $\beta_{0} >
\lambda$, the points $A_{1}^{+}$ and $A_{2}^{-}$ are saddle points, while $%
A_{1}^{-}$ and $A_{2}^{+}$ act as sources. For $\beta_{0} < \lambda$, the
roles reverse: $A_{1}^{+}$ and $A_{2}^{-}$ become sources, while $A_{1}^{-}$
and $A_{2}^{+}$ are saddle points.

\begin{figure}[tbph]
\centering\includegraphics[width=0.5\textwidth]{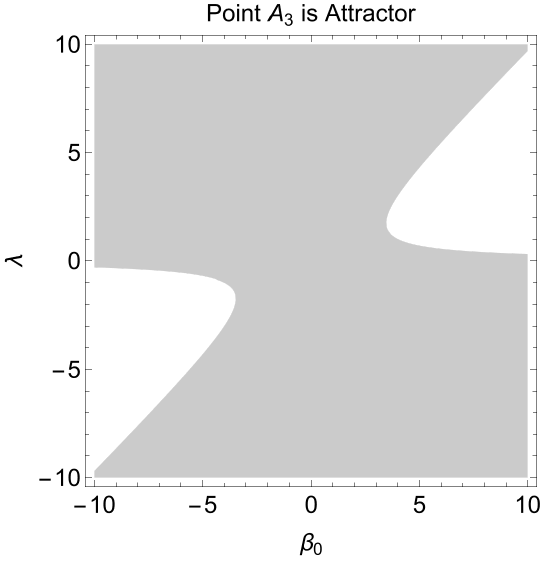}
\caption{Interaction A: Region plots in the space of variables $\left\{ 
\protect\beta _{0},\protect\lambda \right\} $ where $A_{3}$ is an attractor.}
\label{nf1}
\end{figure}

The stationary point $A_{3}$ is well defined for any real value of parameter 
$\lambda $. The physical parameters are derived
\begin{equation*}
\left( \xi ,q\right) _{A_{3}}=\left( 0,-1-\frac{\lambda ^{2}}{2}\right) .
\end{equation*}%
Thus the solution at point $A_{3}$ describes always cosmic acceleration,
driven by the phantom scalar field. The de Sitter universe is recovered when 
$\lambda =0$. The analysis of the linearized system near the stationary
points give that the asymptotic solution is a stable solution and~$A_{3}$ is
an attractor for~$\left\{ \beta _{0}\leq -2\sqrt{3}:~\left( \lambda -\frac{%
\beta _{0}}{2}\right) ^{2}>\beta _{0}^{2}-12\right\} $,$~\left\{ \beta
_{0}^{2}<12\right\} $ or $\left\{ \beta _{0}>2\sqrt{3}:\left( \lambda -\frac{%
\beta _{0}}{2}\right) ^{2}<\beta _{0}^{2}-12\right\} $. In Fig. \ref{nf1} we
present the region in the space of variables $\left\{ \beta _{0},\lambda
\right\} $, in which point $A_{3}$ is an attractor.

\begin{figure}[tbph]
\centering\includegraphics[width=1\textwidth]{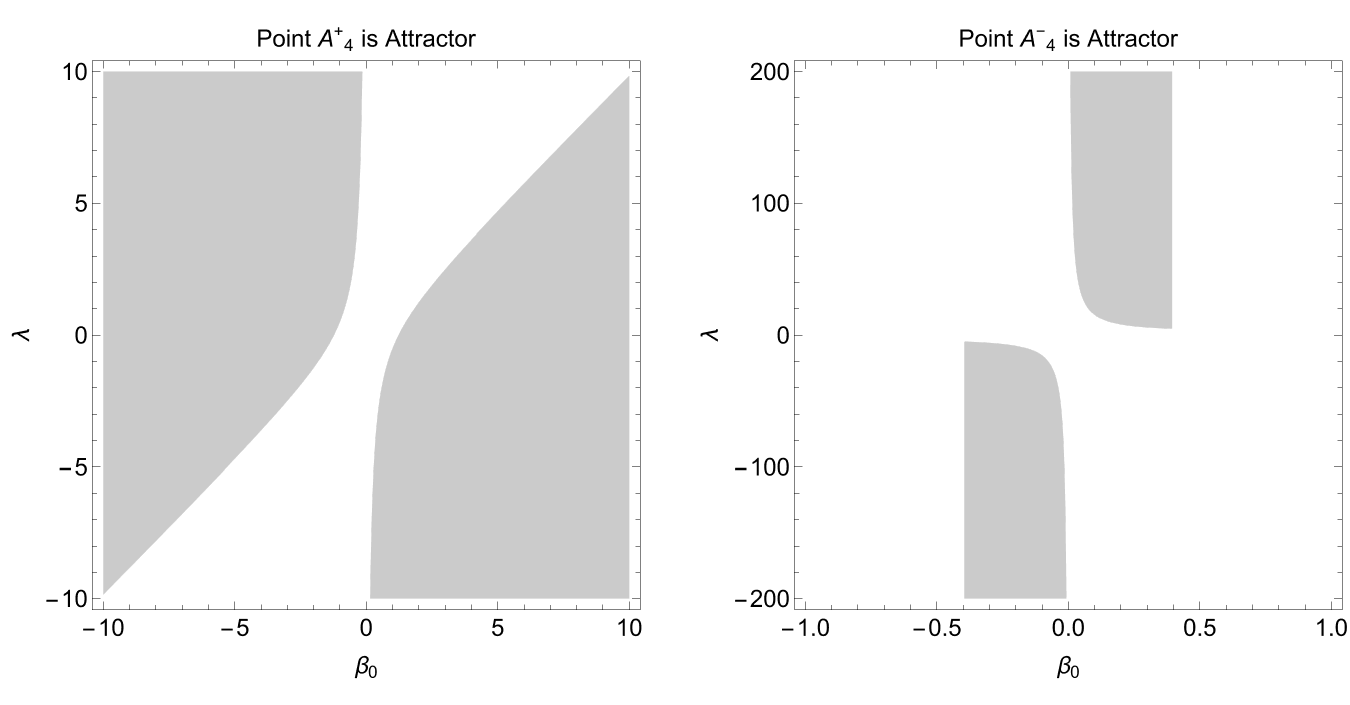}
\caption{Interaction A: Region plots in the space of variables $\left\{ 
\protect\beta _{0},\protect\lambda \right\} $ where $A_{4}^{+}$ is attractor
(Left Fig.) and $A_{5}^{-}$ is attractor (Righ Fig.)}
\label{nf1a}
\end{figure}

The set of point $A_{4}^{\pm }$ are well defined for real values of
parameter $\beta _{0}$. We calculate 
\begin{equation*}
\left\{ \xi ,q\right\} _{A_{4}^{\pm }}=\left( \sqrt{\frac{3+2\beta _{0}^{2}}{%
3+4\beta _{0}^{2}}},\frac{1}{2}-\beta _{0}^{2}\right) .
\end{equation*}%
The asymptotic solution describes a universe with nonzero interacting terms,
in which the two fluids contribute to cosmic evolution. The corresponding
scale factor is given by a scaling function $\beta _{0}^{2}\neq \frac{1}{2}$%
, and from an exponential function, that is, it describes the de Sitter
solution for $\beta _{0}^{2}=\frac{1}{2}$. Last but not least, acceleration
is occurred for $\beta _{0}^{2}>\frac{1}{2}$. The eigenvalues of the
linearized system gives that point $A_{4}^{+}$ is an attractor for $\left\{
\beta _{0}>0:\lambda >\frac{2\beta _{0}^{2}-3}{2\beta _{0}},\lambda <\frac{%
2\beta _{0}^{2}-3}{2\beta _{0}}\right\} $, while $A_{4}^{-}$ is an attractor
for $\beta _{0}^{2}<\frac{1}{2}\sqrt{\frac{1}{2}\left( \sqrt{105}-9\right) }$%
and $\frac{9+30\beta _{0}^{2}+8\beta _{0}^{4}}{\left( 3+4\beta _{0}\right) }%
<2\beta _{0}\lambda $. The regions are presented in Fig. \ref{nf1a}.

\begin{figure}[tbph]
\centering\includegraphics[width=1\textwidth]{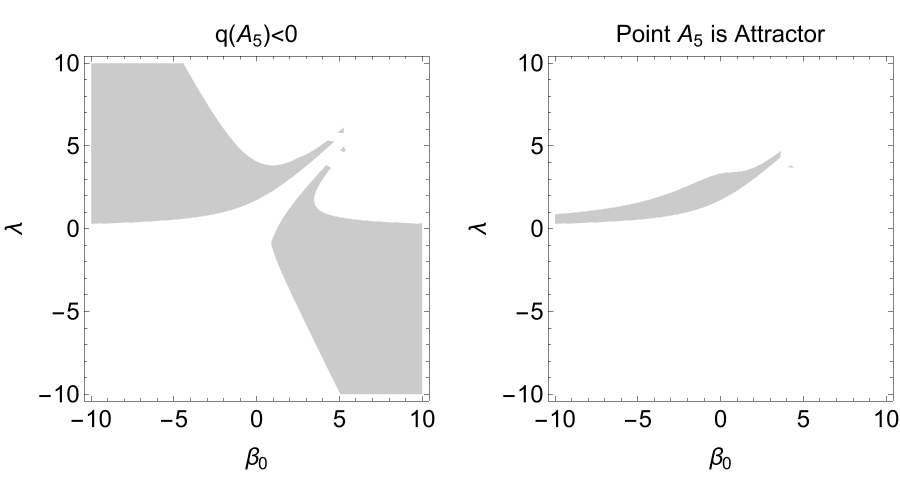}
\caption{Interaction A: Region plots in the space of variables $\left\{ 
\protect\beta _{0},\protect\lambda \right\} $ where $A_{5}$ describes cosmic
acceleration (Left Fig.) and $A_{5}$ is an attractor (Right Fig.).}
\label{nf1e}
\end{figure}

The stationary point $A_{5}$ is well defined for $\left\{ \lambda <0:~\alpha
>0\right\} $, or \newline $\left\{ 0<\lambda \leq 2\sqrt{6}:\alpha \leq \frac{\lambda
^{2}-3}{\lambda },\lambda <\alpha \leq \frac{\lambda ^{2}+3}{\lambda }%
\right\} $ or \newline $\left\{ \lambda >2\sqrt{6}:\alpha \leq \frac{\lambda -\sqrt{%
\lambda ^{2}-24}}{2},\frac{\lambda +\sqrt{\lambda ^{2}-24}}{2}\leq \alpha
\leq \frac{\lambda ^{2}-3}{\lambda },\lambda <\alpha \leq \frac{\lambda
^{2}+3}{\lambda }\right\} $. The stationary points describe interaction in
the dark sectior with 
\begin{equation*}
\left\{ \xi ,q\right\} _{A_{5}^{\pm }}=\left( \sqrt{1-\chi ^{2}\left(
A_{5}\right) -\zeta ^{2}\left( A_{5}\right) },\frac{18 \text{sign}\left( \alpha
-\lambda \right) -15-2\left( \alpha -\lambda \right) \left( 2\alpha +\lambda
\right) }{2\left( 3+2\left( \alpha -\lambda \right) ^{2}\right) }\right) .
\end{equation*}%
Acceleration is occurred for values of the free parameters $\lambda ,\beta
_{0}$ as they are presented in Fig. \ref{nf1e}. Furthermore, the region
space where $A_{5}$ is an attractor is given in Fig. \ref{nf1e}

\begin{figure}[tbph]
\centering\includegraphics[width=1\textwidth]{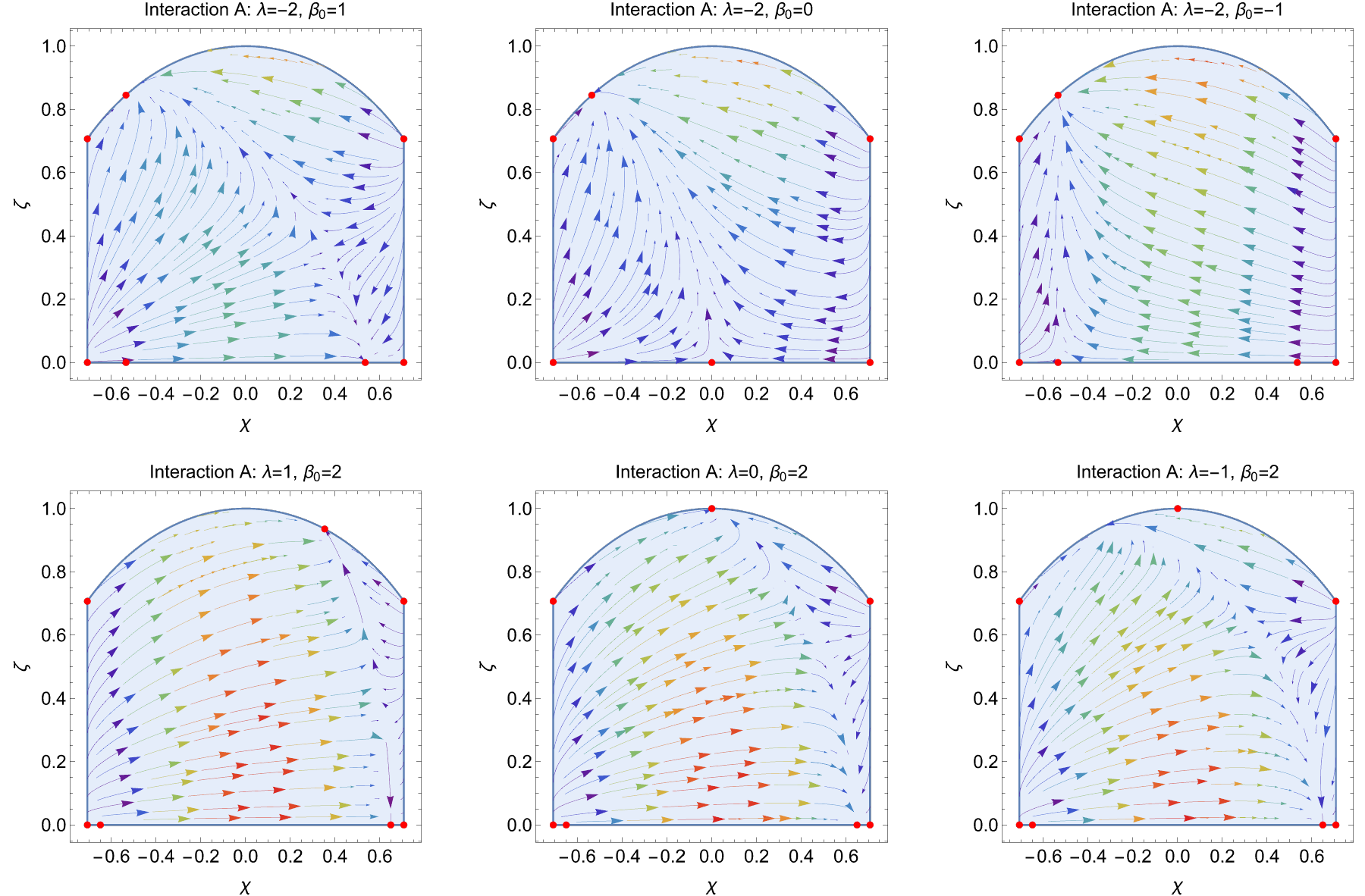}
\caption{Interaction A: Phase-space portraits for the dynamical system (%
\protect\ref{cc.10}), (\protect\ref{cc.11}) for different values of the free
parameters. With red are marked the stationary points. We present the case
where there not any interaction, $\protect\alpha =0$, and the potential
function is constant, $\protect\lambda =0$. }
\label{nf1b}
\end{figure}

\begin{figure}[tbph]
\centering\includegraphics[width=1\textwidth]{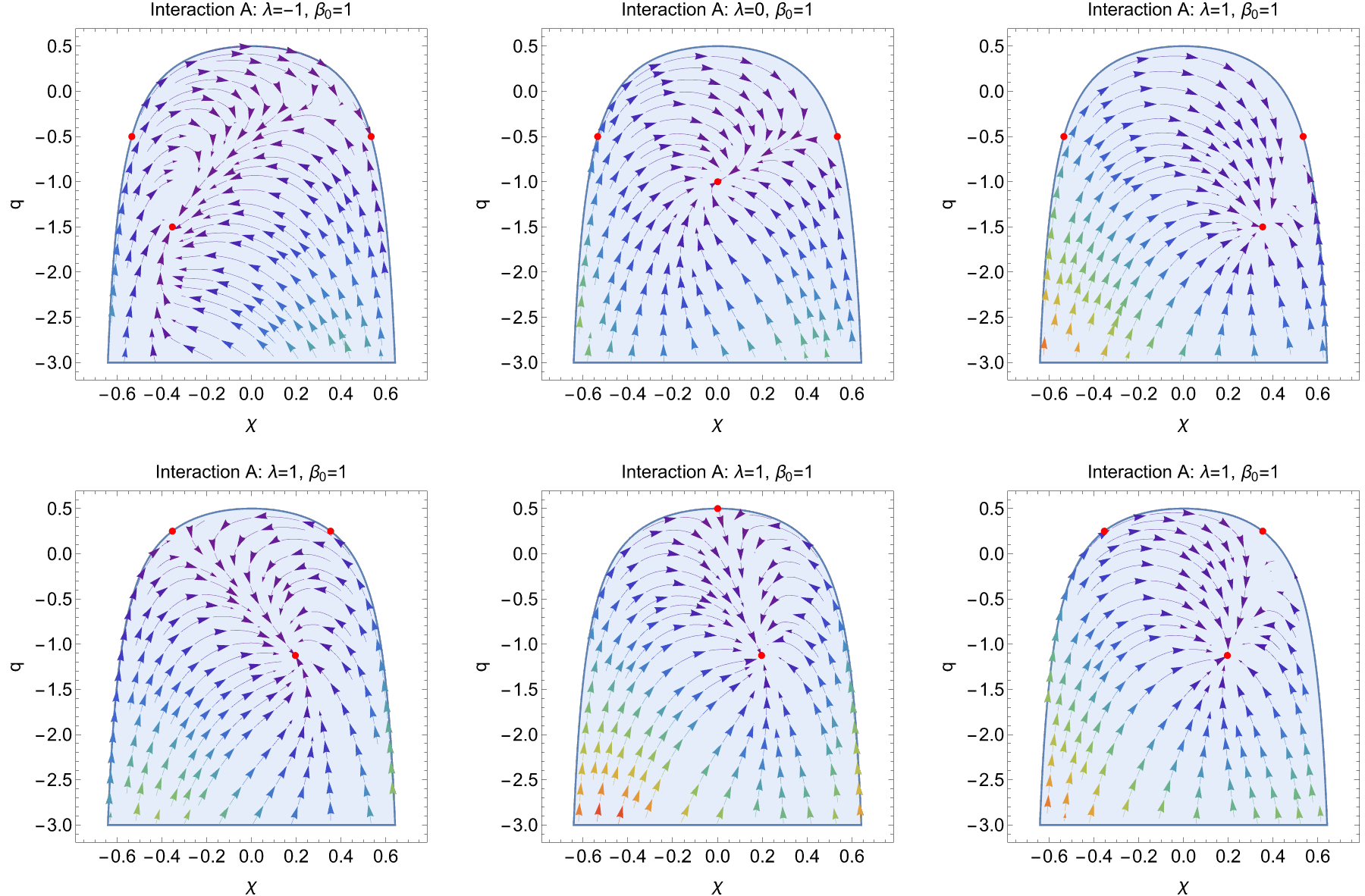}
\caption{Interaction A: Phase-space portraits for the dynamical system (%
\protect\ref{cc.10}), (\protect\ref{cc.11}) for different values of the free
parameters where we have substitute $\protect\zeta ~$from expression (\protect
\ref{cc.08b}) such that the vertical axis to describe the deceleration
parameter~$q$. Color red marks the stationary points at the finite
regime. We present the case where there is not any interaction, $\protect\alpha %
=0$, and the potential function is constant, $\protect\lambda =0$. }
\label{nf1c}
\end{figure}

The above results are summarized in Table \ref{tab1}. Moreover, in Fig. \ref%
{nf1b} we present phase-space portraits for the dynamical system (\ref{cc.10}%
), (\ref{cc.11}) for different values of the free parameters. Finally, in
Fig. \ref{nf1c} we present phase-space portraits in the plane $\left\{ \chi
,q\right\} $.

%TCIMACRO{\TeXButton{B}{\begin{table}[tbp] \centering}}%
%BeginExpansion
\begin{table}[tbp] \centering%
%EndExpansion
\caption{Interaction A: Stationary points and physical propertes}%
\begin{tabular}{ccccc}
\hline\hline
\textbf{Point} & $\mathbf{\xi }\neq 0$ & \textbf{Interaction?} & \textbf{%
Acceleration?} & \textbf{Attractor?} \\ \hline
$A_{1}^{\pm }$ & False & False & Big Rip & False \\ 
$A_{2}^{\pm }$ & False & False & Big Rip & False \\ 
$A_{3}$ & False & False & Always & Fig. \ref{nf1} \\ 
$A_{4}^{\pm }$ & True & True & $\beta _{0}^{2}>\frac{1}{2}$ & Fig. \ref{nf1a}
\\ 
$A$ & True & True & Fig. \ref{nf1e} & Fig. \ref{nf1a} \\ \hline\hline
\end{tabular}%
\label{tab1}%
%TCIMACRO{\TeXButton{E}{\end{table}}}%
%BeginExpansion
\end{table}%
%EndExpansion

\subsection{Interacting Model $Q_{B}$}

For the second interacting model $Q_{B}$, the gravitational field equations
for the exponential potential $V\left( \phi \right) =V_{0}e^{\lambda \phi }$%
, after the use of the constraint equations (\ref{cc.03}), (\ref{cc.03a})
become%
\begin{small}
\begin{align}
\frac{d\chi }{d\tau } =&\frac{1}{2}\left( 2\left( \beta _{0}+\lambda
\right) \zeta ^{2}\left( 1-2\chi ^{2}\right) -\chi \left( \left( 1-\chi
^{2}+\zeta ^{2}\right) \sqrt{6\left( 1-2\chi ^{2}\right) }+2\beta _{0}\chi
\left( 1-2\chi ^{2}\right) \right) \right) ,  \label{cc.20} \\
\frac{d\zeta }{d\tau } =&\frac{\zeta }{2\sqrt{2}}\left( 2\left( \lambda
-\beta _{0}\left( \beta _{0}+\lambda \right) \zeta ^{2}\right) \chi +4\beta
_{0}\chi ^{3}+\left( 1+\chi ^{2}-\zeta ^{2}\right) \sqrt{6\left( 1-2\chi
^{2}\right) }\right) .  \label{cc.21}
\end{align}%
\end{small}
Similar to before, the dynamical variables are constraints as $\chi ^{2}\leq 
\frac{1}{2}$, and $0\leq \zeta \leq 1$, with $\chi ^{2}+\zeta ^{2}\leq 1$.

We proceed with the presentation of the stationary points $B=\left( \chi
\left( B\right) ,\zeta \left( B\right) \right) $ for the dynamical system (%
\ref{cc.20}), (\ref{cc.21}). Due to the algebraic difficulty of the equations, we
present the analysis for the case of $\lambda =0$, that is, the potential
functions are constant.

\begin{eqnarray*}
B_{1}^{\pm } &=&\frac{1}{\sqrt{2}}\left( \pm 1,0\right) , \\
B_{2}^{\pm } &=&\frac{1}{\sqrt{2}}\left( \pm 1,1\right) , \\
B_{3} &=&\left( 0,0\right) , \\
B_{4} &=&\left( -\sqrt{\frac{3}{\beta _{0}^{2}+3+\sqrt{\beta _{0}^{2}\left(
\beta _{0}^{2}-6\right) }}},0\right) , \\
B_{5} &=&\left( -\sqrt{\frac{\beta _{0}^{2}+3+\sqrt{\beta _{0}^{2}\left(
\beta _{0}^{2}-6\right) }}{3+4\beta _{0}^{2}}},0\right) , \\
B_{6} &=&\frac{1}{2}\left( -\sqrt{1+\frac{1}{\sqrt{1+\frac{4}{3}\beta
_{0}^{2}}}},\sqrt{1-\frac{3}{\sqrt{1+\frac{4}{3}\beta _{0}^{2}}}}\right) , \\
B_{7} &=&\frac{1}{2}\left( \sqrt{1-\frac{1}{\sqrt{1+\frac{4}{3}\beta _{0}^{2}%
}}},\sqrt{1+\frac{3}{\sqrt{1+\frac{4}{3}\beta _{0}^{2}}}}\right) .
\end{eqnarray*}

Stationary points $B_{1}^{\pm }$ and $B_{2}^{\pm }$ have the same physical
properties with the points $A_{1}^{\pm }$ and $A_{2}^{\pm }$ respectively.
Furthermore, the stability properties are the same. Indeed, for $\beta
_{0}>0 $, $B_{1}^{+}$, $B_{2}^{-}$ are saddle points, and points $B_{1}^{-}$%
, $B_{2}^{+}$ are sources; while for $\beta _{0}<0$, points $B_{1}^{+}$, $%
B_{2}^{-}$ are sources, and $B_{1}^{-}$, $B_{2}^{+}$ are saddle points.

Point $B_{3}$ describes the matter-dominated epoch, where only the dark
matter contributes to the universe, that is, $\left( \xi,q\right)
_{B_{3}}=\left( 1,\frac{1}{2}\right) $. The eigenvalues of the linearized
system are calculated $-\frac{\sqrt{3}}{2},+\frac{\sqrt{3}}{2}$, from where
we infer that the $B_{3}$ is always a saddle point.

Moreover, points $B_{4}$ and $B_{5}$ describe universes where the kinetic
term of the phantom field interacts with the dark matter. The points are well
defined for $\beta _{0}^{2}\geq 6$, and the physical parameters are 
\begin{equation*}
\left( \xi ,q\right) _{B_{4}}=\left( \sqrt{1-\frac{3}{\beta _{0}^{2}+3+\sqrt{%
\beta _{0}^{2}\left( \beta _{0}^{2}-6\right) }}},\frac{1}{2}\left( 4-\beta
_{0}^{2}+\sqrt{\beta _{0}^{2}\left( \beta _{0}^{2}-6\right) }\right) \right)
,
\end{equation*}%
and 
\begin{equation*}
\left( \xi ,q\right) _{B_{5}}=\left( \sqrt{1-\frac{3}{\beta _{0}^{2}+3+\sqrt{%
\beta _{0}^{2}\left( \beta _{0}^{2}-6\right) }}},\frac{1}{2}\left( 4-\beta
_{0}^{2}-\sqrt{\beta _{0}^{2}\left( \beta _{0}^{2}-6\right) }\right) \right)
.
\end{equation*}%
Thus, point $q\left( B_{4}\right) <0$, for $6\leq a^{2}<8$, while $q\left(
B_{5}\right) <0$ for $\beta _{0}^{2}\geq 6$. Finally, the analysis of the
eigenvalues near these two stationary points give that $B_{4}$ describe
always an unstable solution, while $B_{5}$ is an attractor for $\beta _{0}>%
\sqrt{6}$.

\begin{figure}[tbph]
\centering\includegraphics[width=1\textwidth]{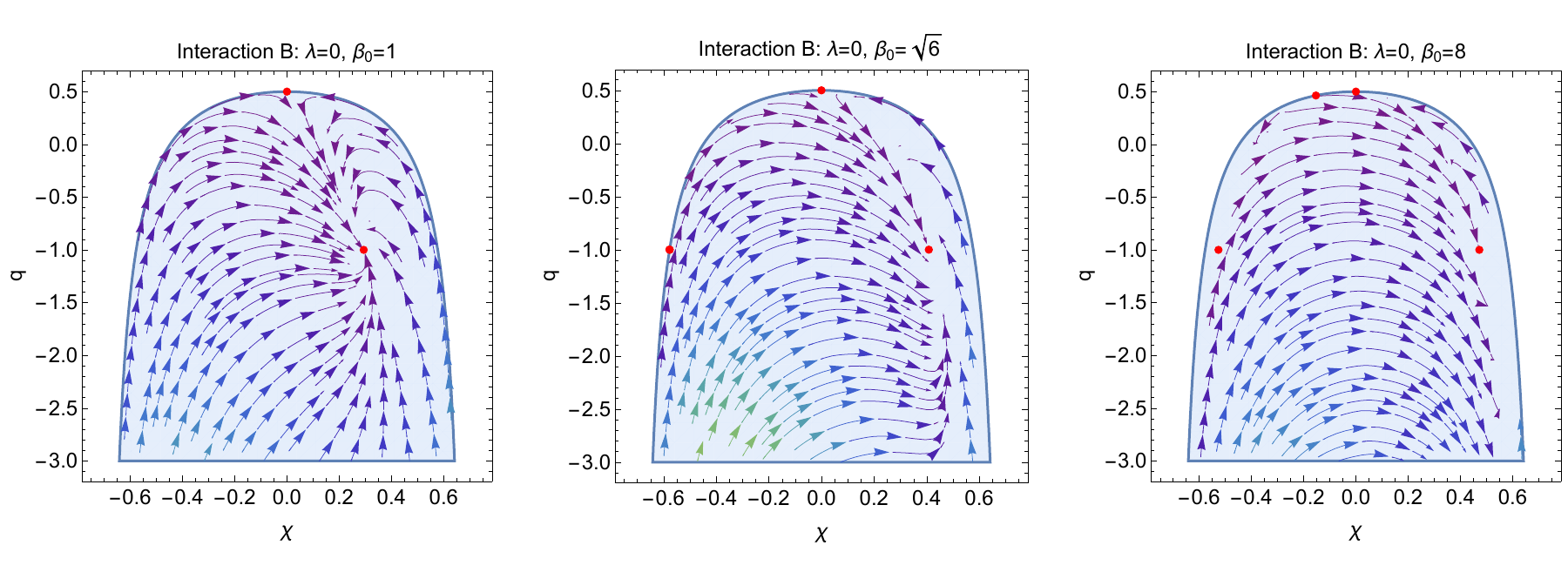}
\caption{Interaction B: Real components for the two eigenvalues for the
linearized dynamical system (\protect\ref{cc.20}), (\protect\ref{cc.21})
near point $B_{7}$. }
\label{nf2}
\end{figure}

The stationary points $B_{6}$ and $B_{7}$ describe\ de Sitter solutions in
which the dark matter interacts with all the components of the scalar field.
Point $B_{6}$ is well defined for $\beta _{0}^{2}\geq 6$. The physical
parameters are calculates%
\begin{equation*}
\left( \xi ,q\right) _{B_{6}}=\left( \sqrt{\frac{1}{2}+\frac{1}{2\sqrt{1+%
\frac{4}{3}\beta _{0}^{2}}}},-1\right) ,
\end{equation*}%
and%
\begin{equation*}
\left( \xi ,q\right) _{B_{7}}=\left( \sqrt{\frac{1}{2}+\frac{1}{2\sqrt{1+%
\frac{4}{3}\beta _{0}^{2}}}},-1\right) ,
\end{equation*}%
As far as the stability is concerned, point $B_{6}$ always unstable, while $%
B_{7}$ is an attractor for the range of parameter $\beta _{0}$ as given in
Fig. \ref{nf2}. 
\begin{figure}[tbph]
\centering\includegraphics[width=1\textwidth]{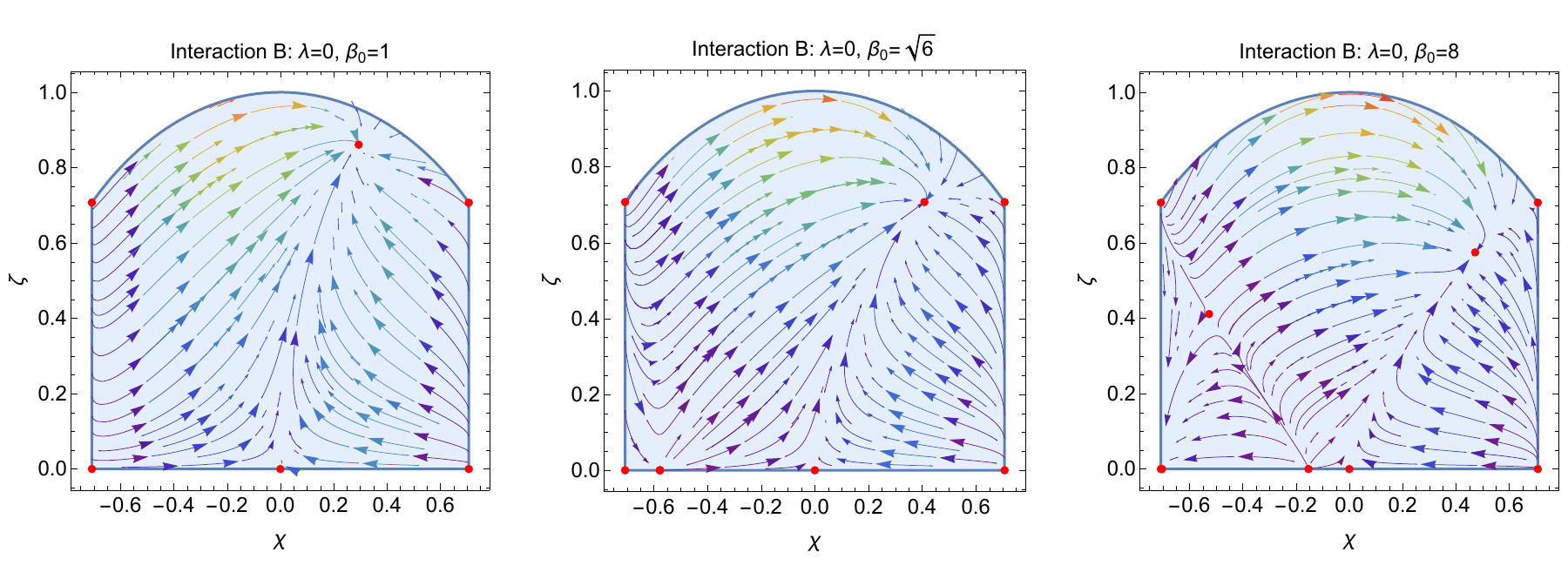}
\caption{Interaction B: Phase-space portraits for the dynamical system (%
\protect\ref{cc.20}), (\protect\ref{cc.21}) for different values of the free
parameters. With red are marked the stationary points. The plots are for $%
\protect\lambda =0$. }
\label{nf2b}
\end{figure}
\begin{figure}[tbph]
\centering\includegraphics[width=1\textwidth]{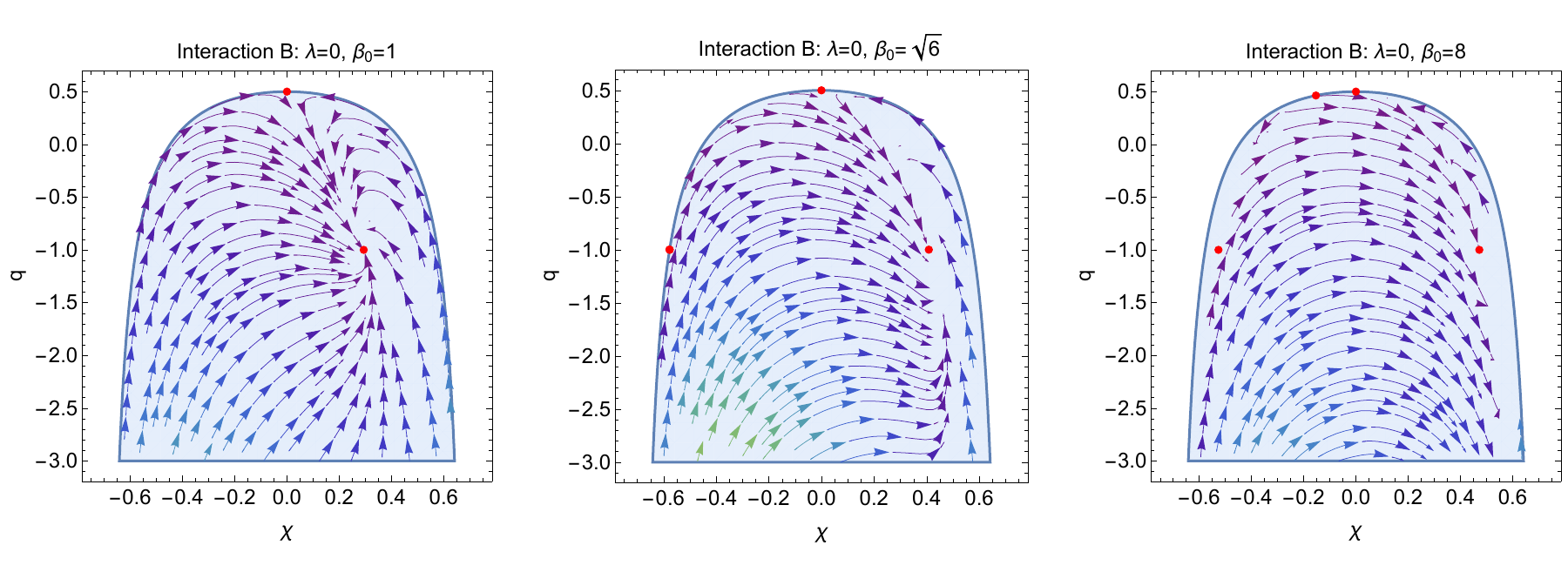}
\caption{Interaction B: Phase-space portraits for the dynamical system (%
\protect\ref{cc.10}), (\protect\ref{cc.11}) for different values of the free
parameters where we have substitute $\protect\zeta ~$from expression (\protect
\ref{cc.08b}) such that the vertical axis to describe the deceleration
parameter~$q$. With red are marked the stationary points at the finite
regime.\ The plots are for $\protect\lambda =0$. }
\label{nf2c}
\end{figure}

The above results are summarized in Table \ref{tab2}. In Fig. \ref{nf2b}, we
present phase-space portraits of the dynamical system (\ref{cc.20}) and (\ref%
{cc.21}) for various values of the free parameters. Additionally, in Fig. %
\ref{nf2c}, we display phase-space portraits in the $\left( \chi, q\right)$
plane.

%TCIMACRO{\TeXButton{B}{\begin{table}[tbp] \centering}}%
%BeginExpansion
\begin{table}[tbp] \centering%
%EndExpansion
\caption{Interaction B: Stationary points and physical propertes}%
\begin{tabular}{ccccc}
\hline\hline
\textbf{Point} & $\mathbf{\xi }\neq 0$ & \textbf{Interaction?} & \textbf{%
Acceleration?} & \textbf{Attractor?} \\ \hline
$B_{1}^{\pm }$ & False & False & Big Rip & False \\ 
$B_{2}^{\pm }$ & False & False & Big Rip & False \\ 
$B_{3}$ & True & False & False & False \\ 
$B_{4}$ & True & True & $6\leq a^{2}<8$ & False \\ 
$B_{5}$ & True & True & True & $\beta _{0}>\sqrt{6}$ \\ 
$B_{6}$ & True & True & True & False \\ 
$B_{7}$ & True & True & True & Fig. \ref{nf2} \\ \hline\hline
\end{tabular}%
\label{tab2}%
%TCIMACRO{\TeXButton{E}{\end{table}}}%
%BeginExpansion
\end{table}%
%EndExpansion

\section{Conclusions}

\label{sec4}

We have considered a spatially flat FLRW geometry where the dark sector of
the universe consists of dark matter, described by a dust fluid source, and
dark energy, modeled by a phantom scalar field. The introduction of the
phantom field allows the equation of state parameter to cross the phantom
divide line and take values smaller than $-1$. We introduced energy transfer
between the components of the dark sector and revised the phase-space
analysis by introducing a novel set of compactified dimensionless variables.

We considered a normalization using the components that constitute the dark
sector of the universe, as defined by expressions (\ref{cc.01}) and (\ref%
{cc.02}). It is easy to see that this normalization is equivalent to the
Hubble normalization in the case of quintessence but differs in the case of
a phantom field. Because the dimensionless variables are compactified, this
approach allows us to formally investigate the phase space and understand
the evolution of the cosmological parameters. As demonstrated, this analysis provides new insights into the phase-space dynamics. 
For the cosmological interacting models, we introduced two cases: model~$A$,
with $Q_{A}\simeq \dot{\phi}\rho _{m}$, and model~$B$, with $Q_{B}\simeq \dot{%
\phi}\rho _{\phi }$. In the first model, the interaction depends on the
energy density of dark matter, while in the second the interaction depends
on the energy density of dark energy. We assumed that the interactions are
proportional to the scalar $\dot{\phi}$, which is necessary to avoid the
appearance of singularities, as discussed recently in \cite{sp2}.

For the first model, we identified five families of asymptotic solutions,
while for the second model, we determined seven families. Big Rip
singularities appear in both models, but they are described by source or
saddle points, indicating that, due to the cosmic interaction, future
singularities are avoided even when a phantom scalar field is introduced.
For both models, the future attractors are scaling solutions. These results
agree with those obtained using Hubble normalization, but are derived here in
a more formal framework.

It is important to mention that for the first model, we assumed the scalar
field potential to be described by an exponential function, while for the
second model, we considered a constant potential. For more general
potentials, the physical properties of the asymptotic solutions change
slightly, as discussed in detail in \cite{sp2}. For this reason, we omit the
presentation of the analysis for other nonlinear scalar field potentials.

In future work, we plan to extend this normalization approach to other
gravitational models, particularly multiscalar field theories.

\begin{acknowledgments}
AP\ \&\ GL thanks the support of VRIDT through Resoluci\'{o}n VRIDT No.
096/2022 and Resoluci\'{o}n VRIDT No. 098/2022. This study was supported by
FONDECYT 1240514, Etapa 2024.
\end{acknowledgments}

\end{document}